\documentclass[prd,aps, reprint,10pt,nofootinbib,superscriptaddress,preprintnumbers]{revtex4-1}
\usepackage{geometry}                
\usepackage{graphicx}
\usepackage{amssymb}
\usepackage{mathtools}
\usepackage{epstopdf}
\usepackage{verbatim}
\usepackage{color}
\usepackage{slashed}
\usepackage{bbold}
\usepackage{fullpage}
\usepackage[title]{appendix}
\usepackage[colorlinks, citecolor=red, linkcolor=blue]{hyperref}

\newcommand{\Lag}{\mathcal{L}}

\begin{document}

\preprint{MCTP-16-10}

\title{\hspace{0.1cm} \\
The lightest visible-sector supersymmetric particle is likely to be unstable}

\author{Bobby~S.~Acharya}
\affiliation{Department of Physics, King's College London, London, WC2R 2LS, UK}\affiliation{The Abdus Salam International Centre for Theoretical Physics, Strada Costiera 11, Trieste, Italy} 
\author{Sebastian~A.~R.~Ellis} 
\affiliation{Michigan Center for Theoretical Physics, University of Michigan, Ann Arbor, MI 48109}
\author{Gordon L. Kane}
\affiliation{Michigan Center for Theoretical Physics, University of Michigan, Ann Arbor, MI 48109}
\author{Brent~D.~Nelson}
\affiliation{Department of Physics, Northeastern University, Boston, MA 02115, USA}
\author{Malcolm~J.~Perry}
\affiliation{DAMTP, Centre for Mathematical Sciences, Wilberforce Road, Cambridge CB3 0WA, UK}

\begin{abstract}

We argue, based on typical properties of known solutions of string/$M$-theory, that the lightest supersymmetric particle of the visible sector will not be stable. In other words, dark matter is {\em not} a particle with Standard Model quantum numbers, such as a WIMP. The argument is simple and based on the typical occurrence of a) hidden sectors, b) interactions between the Standard Model (visible) sector and these hidden sectors, and c) the lack of an argument against massive neutral hidden sector particles being lighter than the lightest visible supersymmetric particle. These conclusions do not rely on arguments such as R-parity violation.

\end{abstract}

\maketitle


\noindent {\it \textbf{Introduction}}. The Standard Model (SM) of particle physics has long been known to lack an adequate candidate for dark matter. An oft-repeated virtue of its minimal supersymmetric extension (the MSSM) is that a conserved matter parity, or R-parity, will imply that the lightest supersymmetric particle is stable, and therefore a natural dark matter candidate~\cite{Ellis:1983ew}. Unfortunately, as we will argue here, when the MSSM is embedded in a ultraviolet (UV) complete theory, such as string or $M$-theory, the lightest supersymmetric particle will not reside in a visible sector.

Whatever the particular details of any particular string compactification may be, there are certain results that appear to be generic. Of particular importance to this paper are the existence of hidden sectors. We define a hidden sector as containing states in the low-energy effective theory that are uncharged at tree-level under the SM gauge symmetries, but can be charged under their own symmetry group $G_{H}$.
Compactified string/$M$-theory solutions will generically have hidden sectors, containing, at a minimum, the gauge fields and gauginos associated the various group factors contained in $G_H$, when the UV solution is supersymmetric
\cite{Dienes:1996zr}-\cite{Corti:2012kd}.
Such sectors have already been used for model-building purposes, in particular for the breaking of supersymmetry (SUSY), which takes place in a hidden sector and is then mediated to the visible sector.

Hidden sectors will of course interact with our visible sector via gravitational interactions, but can also have other so-called ``portals" to the visible sector~\cite{Batell:2009di}. We argue that the existence of hidden sectors and portals leads to the conclusion that the lightest supersymmetric particle in the visible sector (LVSP) is likely to be unstable, since there will generically be a lighter particle in one of the many hidden sectors (an LHSP), into which the LVSP will decay via the portal. For this decay to occur, a simple list of conditions is required, and we find that they are all quite typically available, leading to the conclusion that the LVSP decays. The conditions are the following:
\begin{enumerate}
\item There exists a hidden sector. 
\item There exists a portal connecting the visible sector to the hidden sector. 
\item That hidden sector spectrum includes a particle lighter than the LVSP. 
\end{enumerate}
We will concentrate on the kinetic mixing portal~\cite{Holdom:1985ag}, as we find this to be the most generic portal arising from string theory~\cite{Dienes:1996zr}. The existence of other portals would only serve to strengthen our argument.\\


\noindent {\it \textbf{Hidden Sectors in String Theory}}. A typical feature of the hidden sectors in string/$M$ theory are their multitude and their richness. The presence of hidden sectors is not optional, but often {\em required} to ensure the mathematical consistency of the theory. While systematic studies remain rare (see the discussion in~\cite{Kane:2015qea}), the genericity of large hidden sectors, with many small-rank groups, has been demonstrated in several contexts. These include the heterotic string in the free-fermionic approach~\cite{Dienes:2006ut} and in the orbifold limit~\cite{Giedt:2000bi}, weakly-coupled Type~II string theory~\cite{Cvetic:2004ui}, and Gepner models~\cite{Dijkstra:2004cc}. $F$-theory models are known to produce similarly rich hidden sectors~\cite{Taylor:2015ppa}. Finally, let us consider $M$-theory compactified on a manifold with G$_2$ holonomy. Whilst, in this case, we are technically far away from being able to perform systematic surveys of gauge groups, the general picture is expected to be somewhat similar to the $F$-theory results of~\cite{Taylor:2015ppa}. This can also be argued from duality with the heterotic and Type-II string theory. Given that the number of hidden sectors is bounded by the third Betti number, which is typically $\mathcal{O}(100)$ \cite{Joyce,Corti:2012kd}, it is expected that having many hidden sectors will also prove generic in $M$-theory.

It is important to note that the size of the hidden sector gauge group is often much larger than that of the observable sector. We may use the rank of the group $G_H$ (typically a product of non-Abelian and Abelian factors) as a proxy for the richness of the hidden sector. Traditional string model-building has centred upon weakly-coupled heterotic or Type~II solutions, which typically give a rank for $G_H$ that is larger than $G_{SM}$, but roughly comparable in size. 
In recent years, however, the study of strongly-coupled string theory has been put on a much firmer footing, particularly in the context of $F$-theory. Here the expectation is for the rank of $G_H$ to be much larger than $G_{SM}$, perhaps by orders of magnitude\footnote{Similarly large ranks for $G_H$ have been observed for some time in rational conformal field theory constructions~\cite{Dijkstra:2004cc}.} (see, for example, the specific case studied in~\cite{Taylor:2015xtz}). What is more, the notion of ``generic'' has become increasingly precise in these contexts~\cite{Halverson:2015jua}.

In the next section we will consider the phenomenon of kinetic mixing, which requires the presence of Abelian $U(1)$ factors in the hidden sector.
Given the existence of non-Abelian gauge groups in hidden sectors, it is clear that if they are broken, there can be resulting $U(1)$ factors in the hidden sector. The mechanisms for breaking the hidden sector gauge group can be either via Wilson lines, or via radiative breaking at lower energies. The former case is inherently string-theoretic, in that it requires the presence of non-trivial geometry in the compact space. The latter mechanism is familiar from four-dimensional field theory. Furthermore, there can be $U(1)$'s in the four-dimensional effective field theory that do not stem from non-Abelian groups, and have a string-theoretic origin. A well-studied example is the dimensional reduction of Ramond-Ramond (RR) forms on suitable cycles in Type~II theory~\cite{Jockers:2004yj}. 

%
To be phenomenologically relevant, it is necessary that any such $U(1)$ be non-anomalous, for otherwise the gauge boson would receive a mass of order the string scale through the Green-Schwarz mechanism. In open string theories one typically finds that many of the $U(1)$ factors are anomalous. What is more, many $U(1)$s which are non-anomalous may nevertheless acquire a string-scale mass to satisfy higher-dimensional anomaly cancellation conditions~\cite{Ibanez:1998qp}. Yet the effective mass matrix for the collection of $U(1)$s need not have full rank, and indeed generally does not~\cite{Blumenhagen:2005mu}. The same has been observed in heterotic constructions which generalise the structure group of the gauge bundle~\cite{Anderson:2011ns}. 
%

Furthermore, there are many circumstances in which one expects massless $U(1)$s to emerge. The most obvious are cases in which the $U(1)$ arises from the breaking of a non-Abelian group via Wilson line breaking, or through parallel splitting of stacks of D-branes. Abelian factors arising from the zero modes of closed string RR-forms are guaranteed to be massless on Calabi-Yau surfaces, and can obtain masses only in non-K\"ahler backgrounds~\cite{Grimm:2008ed}. Abelian factors supported by $\overline{D3}$-branes cannot participate in the St\"uckelberg mechanism as the necessary axionic fields are projected out by orientifolding~\cite{Abel:2003ue}. All of these arguments imply that one generically expects light $U(1)$s in the effective field theory.\\


\noindent {\it \textbf{The Kinetic Mixing Portal}}. The kinetic mixing portal was first considered in the context of four-dimensional field theory~\cite{Holdom:1985ag}, in which it arises from the existence, and subsequent integrating out of, heavy bi-fundamental fields, charged under both $U(1)$'s. Such states exist typically in open string theories. For instance, if both $U(1)$'s are supported by $D$-branes which are separated in the extra dimensions, as is the case for all supersymmetric Type I, Type IIA and Type IIB models, then there will be massive open strings which stretch between the two $D$-branes, giving rise to massive bi-fundamentals. There are generalisations of this statement in $M$-theory, $F$-theory and the heterotic string as well. 

These bi-fundamentals will lead to a one-loop mixing of the two $U(1)$ symmetries $U(1)_a$ and $U(1)_b$. In the case of interest let $U(1)_a$ correspond to the visible sector $U(1)_Y$, and $U(1)_b$ correspond to a hidden sector $U(1)$. The Lagrangian of the $U(1)$ kinetic sector then reads
\begin{align}
\Lag_{gauge} = -\frac{1}{4}F^{\mu \nu}_a {F_{a}}_{ \mu \nu}  -\frac{1}{4}F^{\mu \nu}_b {F_{b}}_{ \mu \nu} + \frac{\epsilon}{2}F^{\mu \nu}_a {F_{b}}_{ \mu \nu}
\end{align}
where $\epsilon$ parameterises the mixing of the two $U(1)$'s each with field strength tensor $F^{\mu\nu}_i$.

The expected size of $\epsilon$ can be estimated by calculating the two-point polarisation diagram
and is given by
\begin{align}
\epsilon \simeq \frac{g_a g_b }{12\pi^2}Q_a Q_b \log\left(1+\frac{\Delta m_{ab}^2}{M^2}\right)
\label{epsilon}
\end{align}
where $g_{a,b}~(Q_{a,b})$ are the couplings (charges of bi-fundamentals) of $U(1)_{a,b}$, and $\Delta m_{ab}$ is the mass splitting of the bi-fundamental fields charged under both groups and $M$ is the bi-fundamental mass scale. 
Clearly if the $U(1)$'s sit in an unbroken non-Abelian gauge symmetry, $\epsilon = 0$. If the matter spectrum is charged under a non-Abelian gauge symmetry with a $U(1)$ factor then the mass degeneracy of the spectrum would naively cause $\epsilon = 0$ also. However, this degeneracy is not stable against radiative corrections, and mass splittings $\Delta m_{ab}$ are generated, thus rendering $\epsilon$ non-zero.

These bi-fundamentals may have masses $M \sim {R \over l_s^2} $, where $R$ is the separation of two stacks of $D_p$ branes connected by the open string. This suggests that the mass should be $M \sim \mathcal{O}(M_{GUT})$. Depending on the size of $\Delta m_{ab}$, $\epsilon$ can take on a wide range of possible values. In particular, if the hidden sector gauge group is broken at a scale $M_{G_H} \sim M$ via a Wilson line, $\epsilon$ can be of $\mathcal{O}( 10^{-3})$ for $\mathcal{O}(1)$ charges $Q_a$ and $Q_b$ and $g_a \sim g_b \sim g_Y$. 
On the other hand if the hidden sector gauge group is broken radiatively through field theory dynamics at some much smaller scale  $M_{G_H} \ll M$, $\epsilon$ can be as small as $\mathcal{O}( 10^{-26})$ (e.g. if $M_{G_H} \sim 1$ TeV). Crucially however, barring some non-generic external mechanism to prevent $\epsilon$ from being generated, it is always non-zero~\cite{Dienes:1996zr}. Since $\epsilon$ enters via a dimension-4 operator, it is not suppressed by high mass scales. 

Explicit calculations of the kinetic mixing parameter~(\ref{epsilon}) in Type~II constructions support these arguments. Typical values for the mixing parameter are found to generally lie in the range $10^{-3} \leq \epsilon \leq 10^{-1}$, with values as low as $\epsilon \sim 10^{-6}$ accessible via tuning~\cite{Abel:2008ai}. Some additional volume suppression in the Type~IIB context can be obtained in various LARGE volume scenarios \cite{LVS}, in which the assumption that $g_b \sim g_Y$ is relaxed. In this case, a compact volume which generates an intermediate string scale $M_s \sim 10^{10} \, {\rm GeV}$ could produce an effective mixing parameter in the range $10^{-8} \leq \epsilon \leq 10^{-6}$~\cite{Goodsell:2009xc}. It is unclear whether such large volumes are generic, even within the context of flux compactifications of Type~IIB string theory, though as we will see below, these values still imply that the LVSP will not be an adequate dark matter candidate.

Finally, we should emphasize that non-vanishing kinetic mixing has also been demonstrated in heterotic contexts, including heterotic $M$-theory~\cite{Lukas:1999nh}, Calabi-Yau compactifications~\cite{Blumenhagen:2005ga} and in certain heterotic orbifold limits~\cite{Goodsell:2011wn}. The genericity of kinetic mixing in string theory, and its typical size ($\epsilon \sim 10^{-3}$) appear to bear out the intuition of Dienes et al. from twenty years ago~\cite{Dienes:1996zr}.\\


\noindent {\it \textbf{The Decay Mode}}. Given the existence of a portal, such as the kinetic mixing portal described above, the stability of the LVSP becomes a simple question of kinematics. Note that the usual argument for LVSP stability is based on discrete symmetries, but that requires that the LVSP mass is sufficiently small compared to other particle masses. The LVSP will not decay if it is lighter than all possible combinations of potential hidden sector decay products permitted by gauge invariance alone. One can ask for a sufficient condition for LVSP stability: {\it ‘’why should the LVSP be lighter than all hidden sector particles?’’} A key point is that this question has no obvious answer, and clearly becomes more and more difficult as one increases the number and complexity of the hidden sectors. If there is no good reason for the LVSP to be light compared to hidden sector particles then, presumably, the LVSP will decay. How does it decay?

With supersymmetry the hidden sector contains the $U(1)$ gauge boson and the associated gaugino. The sector may also contain matter charged under the hidden $U(1)$. If any of these states are lighter than the LVSP, then the LVSP can decay via the portal. 
Let us assume that the LVSP is a neutralino, as is common within the MSSM. If a kinetic mixing portal exists to a hidden sector in which the LHSP is also a gaugino, then LVSP decay is expected whenever $\delta m = m_{LVSP}- m_{LHSP} >0$. 
If $\delta m >m_Z$, the neutralino LVSP undergoes 2-body decay to a Z boson with lifetime
\begin{align}
\tau^{\chi_i \rightarrow Z\ \chi_j}_{\rm 2-body} \sim 10^{-17} ~ s \times \left( \frac{10^{-3}}{\epsilon} \right)^2 \left( \frac{0.01}{|N_{i3}N_{j3}^*|}\right)^{2} \ ,
\end{align}
where $N_{km}$ is a neutralino mixing matrix element. We have assumed a mostly Bino or Wino LVSP, and have taken $m_{LVSP}=1$ TeV and $m_{LHSP} =100$ GeV for illustrative purposes. Three-body decays can occur if $\delta m < m_Z$, and may dominate \cite{Arvanitaki:2009hb}. Then the characteristic lifetime is 
\begin{align}
\tau^{\chi_i \rightarrow Z\ \chi_j}_{\rm 3-body} &\sim 10^{-9}~ \mathrm{s}  \times \left( \frac{10^{-3}}{\epsilon} \right)^2 \left( \frac{0.01}{|N_{i3}N_{j3}^*|}\right)^{2} \ ,
\end{align}
where we have taken $m_{LVSP}=1$ TeV and $m_{LHSP} =950$ GeV for illustrative purposes. There are also both two- and three-body decays to a Higgs boson, with $\tau^{\chi_i \rightarrow h\ \chi_j} \sim \frac{H_{ij}^2}{|N_{i3}N_{j3}^*|^2} \tau^{\chi_i \rightarrow Z\ \chi_j}$, where $H_{ij}$ is the neutralino coupling to higgs bosons. Additionally, if the LVSP is mostly Higgsino, $\tau^{\tilde{H}\rightarrow Z\ \chi_j} \sim|N_{i3}N_{j3}^*| ~ \tau^{\tilde{B},\tilde{W}\rightarrow Z\ \chi_j}$ for both the two- and three-body decays. 

There can also be decays into a chiral fermion LHSP which can be much lighter; we will describe the resultant parameter space in a follow-up paper. \\


\noindent {\it \textbf{Summary}}. In this paper we have put forward arguments that imply that the lightest supersymmetric particle in the visible sector is not, in fact, stable. At the very least it is metastable, though it is far more likely that it undergoes prompt decays into hidden sector states. The components of the argument are simple: (1) there is at least one hidden sector, (2) there is at least one portal connecting it to the visible sector, and (3) there exists matter in that sector which is lighter than the LVSP. We have illustrated these arguments with a kinetic mixing portal, since this appears to be the most generic outcome from string theory, but the presence of additional portals would only strengthen the argument.

Several aspects of our argument have appeared elsewhere in the literature in various forms. Here we have emphasised the generic nature of these components in string/$M$-theory solutions. Given the multitude of hidden sectors in string compactifications, it is quite likely that there exists at least one sector that satisfies these assumptions. Therefore we conclude that the LVSP will decay.  It is the generality of this conclusion that compels us to argue for a paradigm shift in the thinking of phenomenologists when it comes to dark matter. In particular it raises the likelihood that dark matter resides in a hidden sector which might be very difficult to probe. 

We conclude by noting that relegating dark matter to some hidden sector has phenomenological consequences. The resulting lifetime will affect collider signatures; we will return to study these in a follow-up paper. The kinetic mixing portal scenario illustrated may be cosmologically perilous, due to long-range forces and millicharged particles, to disruptions in Big Bang nucleosynthesis, to a relic overabundance that conflicts with the known age of the universe~\cite{Goodsell:2009xc,Ibarra:2008kn,Arvanitaki:2009hb}. 
%
%
None of these challenges negate the conclusion that the LVSP is very likely unstable. It is non-generic to avoid sizeable kinetic mixing and light hidden sector states. Instead, it may turn out that the study of dark matter in string theories will illuminate that corner of the string/M-theory landscape in which our world resides.



\noindent {\it \textbf{Acknowledgements.}} BSA would especially like to thank R. Valandro for discussions on Type IIB examples. We also thank J. Ellis, J. Halverson, A. Pierce, G. Shiu, J. Wells, Y. Zhao and B. Zheng for useful discussions. The work of BSA was supported by the STFC Grant ST/L000326/1. The work of SARE and GLK is supported in part by the U.S. Department of Energy, Office of Science, under grant DE-SC0007859. The work of BN is supported in part by the National Science Foundation, under grant PHY-1314774. The work of MJP is supported in part by STFC.


\end{document}